# Iterative training of robust k-space interpolation networks for improved image reconstruction with limited scan specific training samples

Peter Dawood[1*], Felix Breuer[2], Jannik Stebani[2], Paul R. Burd[4], István Homolya[5], Johannes Oberberger[3], Peter M. Jakob[1] and Martin Blaimer[2]

[1] Department of Physics, University of Würzburg, Würzburg, Germany

[2] Magnetic Resonance and X-ray Imaging Department, Fraunhofer Institute for Integrated Circuits IIS, Division Development Center X-Ray Technology, Würzburg, Germany

[3] Department of Internal Medicine I, University Hospital Würzburg, Würzburg, Germany

[4] Institute for Theoretical Physics and Astrophysics, University of Würzburg, Würzburg, Germany

[5] Brain Imaging Centre, Research Centre for Natural Sciences, Budapest, Hungary

*Correspondence to: Peter Dawood

    Department of Physics, University of Würzburg

    Experimental Physics 5

    Am Hubland, 97074 Würzburg, Germany

    Email: peter.dawood@physik.uni-wuerzburg.de

# Abstract

**Purpose:** To evaluate an iterative learning approach for enhanced performance of Robust Artificial-neural-networks for K-space Interpolation (RAKI), when only a limited amount of training data (auto-calibration signals, ACS) are available for accelerated standard 2D imaging.

**Methods**: In a first step, the RAKI model was optimized for the case of strongly limited training data amount. In the iterative learning approach (termed iterative RAKI), the optimized RAKI model is initially trained using original and augmented ACS obtained from a linear parallel imaging reconstruction. Subsequently, the RAKI convolution filters are refined iteratively using original and augmented ACS extracted from the previous RAKI reconstruction. Evaluation was carried out on 200 retrospectively undersampled in-vivo datasets from the fastMRI neuro database with different contrast settings.

**Results**: For limited training data (18 and 22 ACS lines for R=4 and R=5, respectively), iterative RAKI outperforms standard RAKI by reducing residual artefacts and yields strong noise suppression when compared to standard parallel imaging, underlined by quantitative reconstruction quality metrics. In combination with a phase constraint, further reconstruction improvements can be achieved. Additionally, iterative RAKI shows better performance than both GRAPPA and RAKI in case of pre-scan calibration with varying contrast between training- and undersampled data.

**Conclusion**: The iterative learning approach with RAKI benefits from standard RAKI's well known noise suppression feature but requires less original training data for the accurate reconstruction of standard 2D images thereby improving net acceleration.

K E Y W O R D S Parallel imaging, GRAPPA, RAKI, deep learning, complex-valued machine learning.

## 1. Introduction

Since its invention, magnetic resonance imaging (MRI) has become one of the most widespread clinical diagnostic techniques nowadays. It offers numerous benefits such as the absence of ionizing radiation, non-invasiveness and the capability of showing soft tissue structures. However, MRI acquisitions can be time-consuming and the total scanning time remains a crucial factor. Almost all MRI applications such as dynamic MR angiography, perfusion MRI, or imaging of the cardiac function require accelerated imaging to cover their typical time scales. Strategies to shorten the scan times based on hardware modifications have reached engineering as well as physiological limits, for example due to peripheral nerve stimulations. In order to further decrease scan time, data acquisition techniques based on gradient sub-encoding were considered. Parallel Imaging (PI) is nowadays the most common acceleration strategy in clinical routine. In PI, the MR signal is acquired simultaneously with multiple, independent receiver coils (so-called phased arrays (1) ), while the inverse image space (also known as k-space) is sub-sampled. Dedicated PI reconstruction methods make use of the inherent spatial encoding capabilities of the phased array to recover the full image content. They can be classified to operate either in image- or k-space domain. Image domain methods are essentially based on sensitivity encoding (SENSE) (2) and recover artifact-free images by utilizing explicit spatial coil-sensitivity information. GeneRalized Autocalibrating Partial Parallel Acquisition (GRAPPA) (3) is a widely used method operating in k-space and estimates missing k-space signals by a convolution of adjacent multichannel k-space signals. The convolution filters (also known as GRAPPA kernel) are calibrated by linear least-squares fit using several fully sampled auto-calibration signals (ACS) that serve as scan-specific training data. However, severe noise enhancement due to ill-conditioned matrix systems at high accelerations is a major limitation in all PI methods.

To overcome this limitation, GRAPPA has been generalized recently within the machine learning framework by the deep learning method Robust Artificial-Neural-Networks for k-space interpolation (RAKI) (4). In contrast to GRAPPA, where only one convolution filter layer is applied for k-space interpolation, RAKI exploits multi-layer feature extraction. In RAKI, non-linearity is introduced by applying a non-linear activation function element-wise to the convolution-layers. The combination of multiple convolution layers with non-linear activation functions are essential elements of a convolutional neural network (CNN). Similar to GRAPPA, the neural network parameters in RAKI (i.e. the convolution filter weights within the CNN) are calibrated using scan-specific ACS as training data. In previous studies, RAKI has demonstrated better performance in comparison to GRAPPA (4)(5)(6). However, RAKI requires more training data due to its increased parameter space, which may limit its applicability in standard 2D imaging.

In the field of machine learning, a common way to deal with limited training data includes augmentation strategies (7)(8). For example, images can be rotated, flipped or resized to generate additional training samples. However, applying these augmentation strategies in a straightforward manner does not work for multi-channel k-space interpolation methods such as RAKI (this is elaborated in more detail in the methods section). The goal of k-space interpolation methods is to find an optimal combination (i.e. convolution filters) of measured k-space samples to reconstruct a missing sample. This combination strongly depends on the specific scan-setup including coil geometry, slice orientation and phase-encoding direction(9)(10). For example, the optimal convolution filters are expected to differ between two neighboring imaging slices or between different breathing states as the coil sensitivities have different profiles. Hence, RAKI is often referred to as a scan-specific machine learning approach that does not depend on large databases for training.

This is contrary to many deep-learning parallel imaging reconstructions operating in image-space. They are often related to traditional supervised machine learning and generally rely on CNNs, which can be used to define generalized regularizers. This allows for higher computational complexity compared to regularizers that are used in classical compressed sensing (11), such as image gradient or Tikhonov. The supervised learning process requires a set of fully sampled raw k-space data, acquired in a phased-array coil setup. The data is retrospectively undersampled and forms the input of the reconstruction network. The fully sampled data serves as the target reference image (12). Thus, the reconstruction parameters are learned from a large number of raw datasets.

In this work, we address the issue of limited training samples in RAKI by proposing an iterative training process, termed iterative RAKI (iRAKI). We demonstrate the flexibility of iRAKI for several 2D imaging and reconstruction scenarios with different contrast settings, different calibration approaches (e.g. with pre-scan or integrated ACS) and implementation of phase-constrained reconstruction. Part of this work has been presented at the annual meeting of the ISMRM 2022 (13).

## 2. Methods

### 2.1. Network Optimization

The network architecture used in this work is depicted in supporting information Figure S. 1 A. We implemented a single CNN for simultaneous multi-coil k-space interpolation, rather than assigning each single coil one CNN as implemented in original RAKI. In this way, the correlations and interactions between all coils are preserved. Furthermore, instead of performing real-valued convolution, we implemented its complex-valued equivalent (14)(15). As nonlinear activation function, we used the leaky variant of the complex rectifier linear unit ($\mathbb{C}$ReLU) (15)(16):

$$\mathbb{C}\text{LeakyReLU}(z) = \text{LeakyReLU}(\text{Re}\{z\}) + i\,\text{LeakyReLU}(\text{Im}\{z\}), \quad (1)$$

with Re{$z$} and Im{$z$} denoting the real and imaginary part of signal $z$, respectively, $i$ denoting the imaginary unit, and LeakyReLU is the leaky variant of the standard real-valued rectifier linear unit (17).

Note that the total number of hidden layers, the assigned channel-number for each layer, the learning rate as well as the slope for $\mathbb{C}$LeakyReLU were determined heuristically in a hyperparameter search, as currently there are no existing methods for optimally tuning the network-architecture in deep learning applications (18)(4). Especially the convolution kernel sizes were optimized for the case of a limited training data amount. The hyperparameter optimization resulted in the following network architecture: The input layer $s_1$ takes the complex-valued, zerofilled multi-coil k-space data, resulting in $N_c$ total input channels, with $N_c$ being the number of receiver coils. The hidden layers $s_2$ and $s_3$ are then calculated through linear, complex-valued convolution, and an element-wise activation using a leaky complex Rectifier Linear Unit $\mathbb{C}$LeakyReLU: $s_2 = \mathbb{C}Leaky\text{ReLU}(s_1 \circledast W_\mathbb{C}^1)$ and $s_3 = \mathbb{C}\text{LeakyReLU}(s_2 \circledast W_\mathbb{C}^2)$, with the complex convolution matrix $W_\mathbb{C}^1$ of size $k_y \times k_x = 2 \times 5$ and $W_\mathbb{C}^2$ of size $k_y \times k_x = 1 \times 1$ in phase-encoding and readout direction, respectively. The first hidden layer $s_2$ is assigned 256 channels and the second hidden layer $s_3$ is assigned 128 channels. The output layer $s_4$ predicts all missing points across all coils simultaneously, thus having $(R-1) \times N_c$ channels, where $R$ denotes the undersampling rate. It is activated with the identity function $\gamma(x) = x$, thus reading $s_4 = \gamma(s_3 \circledast W_C^3)$, with $W_C^3$ of size $k_y \times k_x = 1 \times 5$. The mean-squared-error (MSE) of signal prediction $y$ to its groundtruth $\hat{y}$ was used as cost function $L(y, \hat{y})$ for training

$$L(y, \hat{y}) = \frac{1}{N}\left(\sum_{i=0}^{N} |y_i - \hat{y}_i|^2\right), \qquad (2)$$

with $N$ denoting the total number of training samples. The Adaptive Moment Estimation (Adam) optimizer (19) was chosen as optimization algorithm to minimize the mean-squared-error of estimated k-space data to its ground-truth. Bias terms were excluded in the CNN, as they may perturb k-space scaling (4). The CNN was implemented within the PyTorch package 1.8.0 (20). To obtain the final reconstructed image, the interpolated k-spaces are Fourier-transformed and combined by root sum-of-squares.

We compared our RAKI implementation against the publically available original implementation on 200 datasets assembled from the fastMRI neuro database (21) (see section 2.3 for details). The datasets were 4-fold retrospectively undersampled, and 22 ACS lines were used for evaluation.

Note that the GRAPPA k-space reconstruction via convolution can be formulated as $s_{int} = \gamma(s_1 \circledast W_C^G)$, where $s_{int}$ denotes the interpolated k-space signals, $s_1$ denotes the undersampled, multi-coil k-space data, $W_C^G$ is the GRAPPA kernel, and $\gamma$ is the identity function assigned to the only convolution layer. Thus, the model in GRAPPA can be obtained from the RAKI model by omitting the hidden layers, which are used for abstract multi-layer feature extraction of k-space signals. Thus, essentially, GRAPPA can be seen as a reduced version of RAKI (4). All reconstructions were performed on a high-performance-computing-cluster with Intel® Xeon® Gold 6134 (CPU).

## 2.2. Iterative Training

As k-space interpolation is based on correlations between adjacent points and redundancies induced by coil sensitivity profiles, the convolution filter size determines the extent in which the k-space footprint of spatially varying coil sensitivity profiles is captured.

Previous works have shown that a larger kernel size in general is beneficial for k-space interpolation as it yields improved image reconstruction (22). However, the use of a larger kernel requires more ACS. It is worth mentioning that within the general machine learning context, the common strategy to handle the issue of limited training data is to use data augmentation techniques to synthetically enlarge the effective amount of training data. Sandino et al. trained unrolled neural networks on augmented 2D cardiac cine MRI data in image space (23), e.g. by random flipping along readout- and phase encoding direction, or random circular translations along phase-encoding direction. However, these augmentation techniques do not work for k-space interpolation in standard 2D imaging, since the multi-channel k-space correlations must be preserved. The phase-encoding direction must be coherent with the underlying coil geometry for estimation of convolution filters. However, an image rotation, -stretching or –compression comes along with corresponding operations on the coil geometry. Thus, the coherence of the convolution filter and k-space points is lost, as the underlying basis functions to be combined change as well. Thus, one cannot find the searched filter parameters with the augmented data for standard Cartesian undersampling.

Nencka et al. performed RAKI on multi-band imaging, and augmented training data by creating synthetically aliased k-space data via linear combination of the k-space acquisitions for subsets of slices in the excited package (24). For standard 2D imaging, however, this approach is not applicable due to the lack of multiple slice excitations.

As illustrated in Figure 1, the proposed iterative training in iRAKI includes different amounts of original and augmented ACS, as well as different convolution filter sizes(21)(24). The goal is to enhance RAKI image quality when only a reduced amount of original acquired ACS is available. An initial GRAPPA reconstruction is performed in order to obtain augmented ACS (original ACS are re-inserted into reconstructed k-space). In this work, GRAPPA is assigned a kernel size $2 \times 5$. From the initial GRAPPA k-space reconstruction, $N = 65$ central lines are used to train RAKI. The first hidden layer in RAKI is assigned a filter size $4 \times 7$ in PE- and RO-direction, respectively. Inspired by iterative-GRAPPA (25), subsequent iteration steps follow to refine the CNN weights. In each iteration, the CNN weights are transferred, and further optimized using $N' = 65$ central lines (including re-inserted original ACS) from the RAKI reconstruction of the previous iteration. The initial learning rate $\eta_0$ passed to the Adam optimizer is decreased by a constant factor $\Delta\eta$ after each subsequent iteration step. Thus, the learning rate at iteration step number $j$ reads $\eta_j = \eta_0 - j \cdot \Delta\eta$. In this work, we set empirically $\eta_0 = 5e - 3$ and $\Delta\eta = 2e - 4$ for $R = 4$, and $\Delta\eta = 3e - 4$ for $R = 5$. We chose $\Delta\eta$ heuristically such that the cost function (MSE) does not diverge at late iteration steps, ensuring robustness in the iterative training procedure. Accordingly, the total number of iterations $N_{iter}$ amounts to $N_{iter} = \eta_0/\Delta\eta$. The final optimized CNN interpolates the multi-coil sub-sampled k-spaces simultaneously, which are then Fourier-transformed to image domain, and combined via root sum-of-squares coil combination to obtain the final reconstructed image. In the spirit of reproducible research, code to generate results will be made publically available if and when the manuscript is accepted.

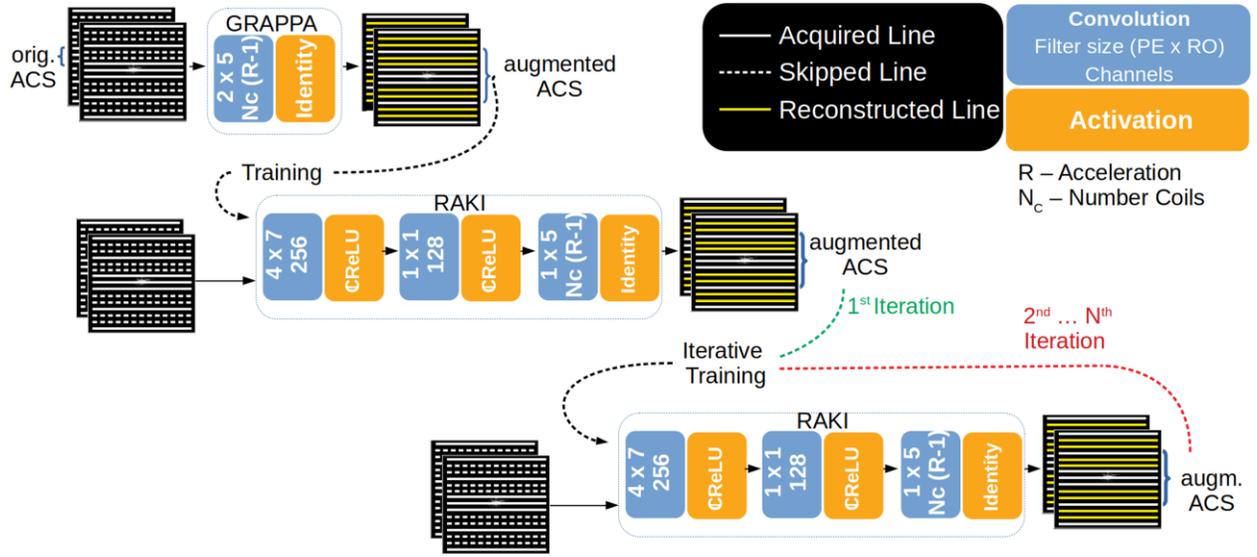

**Figure 1:** Workflow of iRAKI. RAKI is initially trained with augmented ACS obtained from an initial GRAPPA reconstruction using a kernel size $2 \times 5$ (Phase- x Readout-direction). The first hidden layer in RAKI is assigned a convolution filter size $4 \times 7$. From the initial GRAPPA reconstruction, $N = 65$ central k-space lines are used as training samples for RAKI training In subsequent iteration steps, the CNN weights are further optimized using $N' = 65$ central lines from the RAKI reconstruction of the previous iteration step as ACS. Original ACS are inserted after each reconstruction step. The learning rate is decreased by a constant factor after each iteration step, which determines the total iteration number, given an initial learning rate.

## 2.3. Experiments

**Datasets**

To study the performance of iRAKI across a larger cohort, it was tested on 10 fully sampled datasets randomly selected from the fastMRI (21) neuro-database, with the first 5 slices reconstructed in each case. Four different contrasts (T1, T1post, T2 and FLAIR) were considered, resulting in a total of 200 slices for evaluation. For training, 18- and 22 original ACS lines were used for 4-and 5-fold retrospectively undersampling, respectively.

Additionally, three in-plane brain imaging datasets were acquired on healthy volunteers at 3 T (Siemens Magnetom Skyra, Siemens Healthineers, Erlangen, Germany) using a 20-channel head-neck coil array, with only 16 coils activated. The study was approved by our institutional review board. Written informed consent was obtained before each in vivo study.

One dataset is referred to as neuro1, acquired using FLASH with T1-weighting (Rep.-/Echo-Time: 250/2.9 ms, flip angle: 70°, FOV: 230x230 mm$^2$, matrix-size: 320x320, slice-thickness: 3.0 mm).

Furthermore, a T1- and T2-weighted neuro imaging was carried out subsequently (referred to as neuro2) using TSE with the following imaging parameters for T1-weighting: Rep.-/Echo-Time: 500/10.0 ms, flip angle: 90/180°, FOV: 193x220 mm$^2$, matrix-size:224x256, slice-thickness: 4.0 mm. The T2-weighting imaging parameters read: Rep.-/Echo-Time: 4500/102.0 ms, flip angle: 90/180°, FOV: 193x220 mm$^2$, matrix-size: 230x256, slice-thickness: 4.0 mm.

This experiment was carried out on another subject (neuro3) using TSE with a fat saturation module for the T1-weighted image with following imaging parameters: Rep.-/Echo-Time: 600/6.4 ms, flip angle: 90/180, FOV: 199x220 mm, matrix-size:320x320, slice-thickness: 4.0 mm. The T2-imaging parameters read: Rep.-/Echo-Time: 4500/95 ms, flip angle: 90/180, FOV: 200x220 mm, matrix-size: 320x320, slice-thickness: 4.0 mm.

A fourth dataset (referred to as neuro4) was acquired using FLASH with T1-weighting (Rep.-/Echo-Time: 250/3.1 ms, flip angle: 70, FOV: 195x250 mm, matrix-size: 250x320, slice-thickness: 4.0 mm). This scan was prospectively undersampled at rate 4. A pre-scan with PD-weighting (matrixsize: 64x64) was acquired beforehand to study the performance of iRAKI in case of strongly varying contrast information (see section 'Pre-Scan calibration' below).

### In-line calibration

All datasets were retrospectively undersampled, and 18 and 22 ACS lines were used as training data for 4-and 5-fold uniform undersampling, respectively. Original ACS were re-inserted into the final reconstructed k-space. To provide comparable results to image-based deep-learning approaches, both the reference- and the reconstructed image were masked to compute numeric metrics such as normalized mean squared error (NMSE), structural-similarity-index-measure (SSIM) (26) and peak-signal-to-noise-ratio (PSNR). The masks were derived from the standard coil mapping procedure with ESPIRiT (27). In addition, all image reconstructions were evaluated qualitatively via error images.

The model complexity of the CNN in RAKI is significantly determined by its convolution filter sizes. However, analogous to GRAPPA, the choice of the filter sizes affects the total number of available training data, given a fixed number of ACS lines (22). As both complexity and total number of training data are crucial factors for the performance of the CNN, a more detailed evaluation for this trade-off is obligatory. For this purpose, we vary the number of ACS lines between $N_{ACS} = 15, \dots, 100$ with step size 5, and assign two different convolution filter sizes to the first convolution filter in the CNN (denoted as $\boldsymbol{W}_C^1$ in section 2.1): $k_y \times k_x = 2 \times 5$ and $4 \times 7$, respectively. Image reconstruction quality is assessed qualitatively via error images and quantitatively via the normalized mean squared error (NMSE) of the magnitude image w.r.t. the fully sampled reference image. For comparison, the NMSE of the corresponding GRAPPA reconstruction is evaluated. In all cases, in order to investigate the reconstruction performance only, ACS were not re-inserted into the final reconstructed k-space.

### Pre-Scan calibration

In the in-line calibration acquisition scheme, the scan-specific training data used to calibrate the GRAPPA kernel and the model weights in iRAKI can be re-inserted into the reconstructed k-spaces, since the ACS are an integral part of the image scan. Alternatively, the training data can be obtained by acquiring a fully sampled, low resolution pre-scan before the actual undersampled image scan series. As no contrast information is needed in the standard parallel imaging, the pre-scan sequence parameters like repetition- and echo time can be adjusted to maximize SNR, or to minimize the total acquisition time. However, in this case the training data cannot be re-inserted into to the reconstructed k-space of the actual image scan. Thus, the performance of iRAKI needs to be investigated separately for pre-scan ACS with different contrast. For this purpose, we acquired a proton-density weighted pre-scan of size 64 × 64 in PE-and RO-direction, which was used to calibrate GRAPPA, RAKI and iRAKI. The calibrated models were then used to reconstruct a subsequently acquired, 4-fold prospectively undersampled 2D neuro image scan with T1-weighting (referred to as neuro4).

## Phase-constrained reconstruction

The Virtual Conjugate Coils (VCC) concept (28) has been introduced to improve parallel MRI performance by utilizing conjugate symmetry properties of the k-space, and can be seen as a phase-constrained reconstruction technique. From actual physical coils, additional virtual coils are generated which contain conjugate symmetric k-space signals. Thereby, additional image phase and coil phase information is utilized to improve reconstruction conditions. The VCC concept has been presented as a practical approach especially in combination with GRAPPA, since no explicit spatial phase information is required. In this work, we study the influence of VCCs on the reconstruction quality of RAKI and iRAKI in comparison to GRAPPA. The additional k-space signals from a virtual coil can be generated from an actual coil $h$ according to

$$s_{h+N_c}(k) = s_h^*(-k), \quad h = 1 \dots N_c, \tag{3}$$

where $N_c$ denotes the number of actual physical coils in the phased array, $k$ denotes a k-space vector and $s_h^*$ is the complex conjugate signal assigned to coil $h$. The stack of virtually received k-spaces thus contains two times as many coils as actual coils. The reconstruction process is carried out by first generating the virtual coils for both ACS and undersampled data according to Eq. 3, and subsequently performing a k-space reconstruction using a standard GRAPPA-, RAKI- or iRAKI reconstruction. The resulting images of the physical coils are then combined using a root sum-of-squares combination.

**Comparison to end-to-end Variational Network**

We compared the performance of iRAKI to an image-based variational network (VarNet) (29) for two scenarios:

- Single anatomy: VarNet training on knee datasets, image reconstruction of undersampled knee datasets with matching contrast

- Cross-anatomy: VarNet training on knee datasets, image reconstruction of undersampled neuro datasets with different contrast (i.e. scans from the fastMRI cohorts)

We used a variational network from BART (30) that was pre-trained on 20 uniformly undersampled (R=4) knee datasets (10 slices each) with proton-density weighting. The training data was assembled from knee datasets of proton-density weighting of the original VarNet publication (29). Coil sensitivity maps were computed with ESPiRIT (27) using 27 ACS lines. Same undersampled datasets were reconstructed with iRAKI using 27 original ACS lines for training.

## 3. Results

### 3.1. Network optimizations

The optimized RAKI model was compared to the original RAKI implementation on a total of 200 datasets assembled from the fastMRI neuro database. At 4-fold retrospectively undersampling and 22 ACS lines as training data, our optimized variant achieved NMSE-medians reduced by 33.51 %, 43.21 %, 34.53 %, and 27.98 % for T1, T1post, T2 and FLAIR-weighting, respectively (see supporting material Figure S1 B). On a visual scale, the optimized network strongly suppresses residual artifacts compared to the original implementation (see supporting material Figure S1 C). Based on these results, we consider the optimized variant as the favored model for this manuscript, and all subsequently shown RAKI-reconstructions were obtained from the RAKI implementation described in the methods section 2.1. and denoted as standard RAKI.

## 3.2. Experiments

**Training data amount**

Figure 2 depicts the NMSE of GRAPPA, RAKI and iRAKI at 4-fold retrospectively undersampling of the neuro1 dataset in dependence of the number of ACS lines in range 15-100. Both GRAPPA and RAKI were evaluated with two different kernel sizes (2x5 and 4x7 in phase-encoding (PE) x readout (RO) direction).

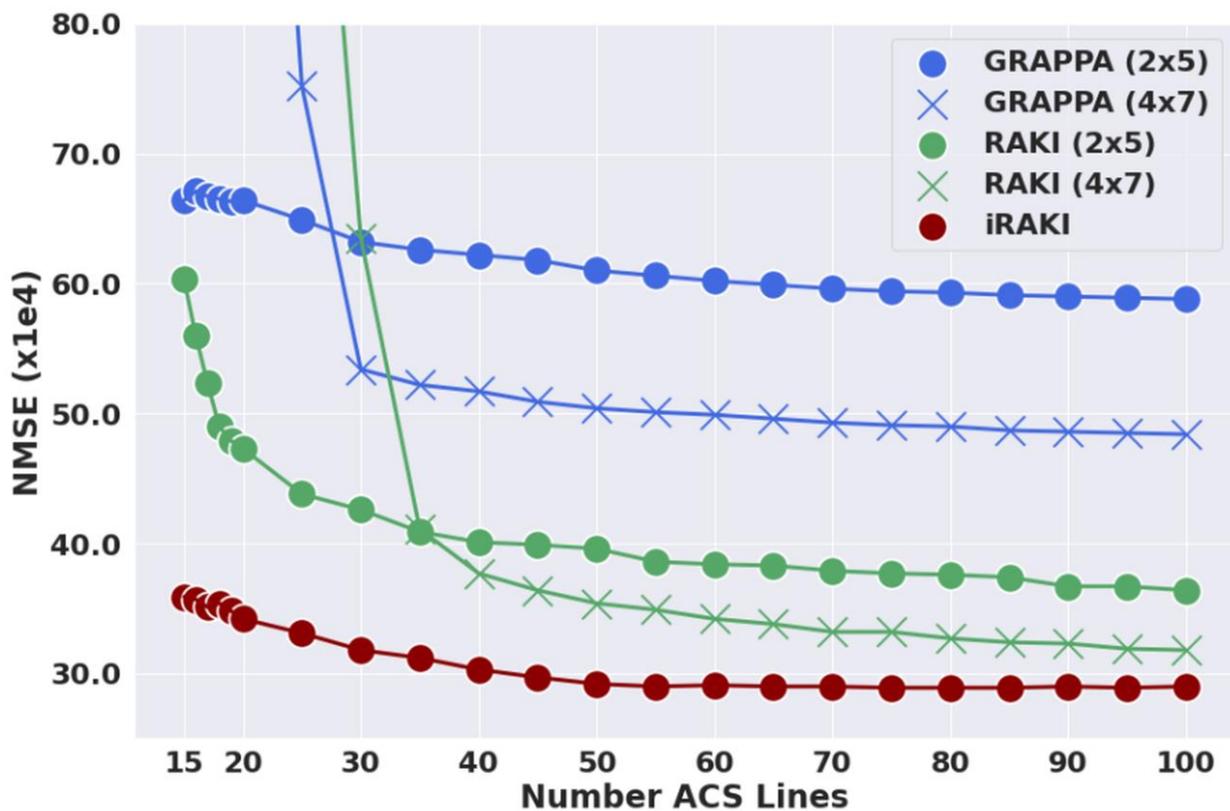

**Figure 2:** The normalized mean-squared-error (NMSE) of GRAPPA, standard RAKI and iRAKI in dependence of the number of ACS lines used as training data for 4-fold undersampled neuro1 dataset. GRAPPA and standard RAKI were evaluated for two kernel sizes assigned to the first convolution layer (2x5 and 4x7 in phase-encoding x readout direction). ACS were not re-inserted into reconstructed k-spaces. Corresponding image reconstructions are included in supporting information Figure S2.

While RAKI deteriorates heavily for ACS amounts less than 20 lines at both kernel sizes, we notice that iRAKI is much more robust in the limit of only a few ACS lines (note that training data is not re-inserted). For ACS amounts exceeding 50 lines, the larger kernel size in standard RAKI outperforms the smaller kernel size, which is the reason why the former is favored in iRAKI. On a visual scale (see Figure 3), we notice that iRAKI trained with 15 original ACS lines yields a reconstruction quality that is comparable to standard RAKI trained with 100 original ACS lines (also shown by numeric results from top to bottom: NRMSE, PSNR and SSIM). For all configurations evaluated in Figure 2, corresponding image reconstructions with error maps and quantitative metrics have been compiled in a movie, which can be found in supporting information Figure S2. The movie underlines that residual artifacts, which appear in standard RAKI at low ACS amounts (<25 lines) are being gradually suppressed with successively increasing ACS amount.

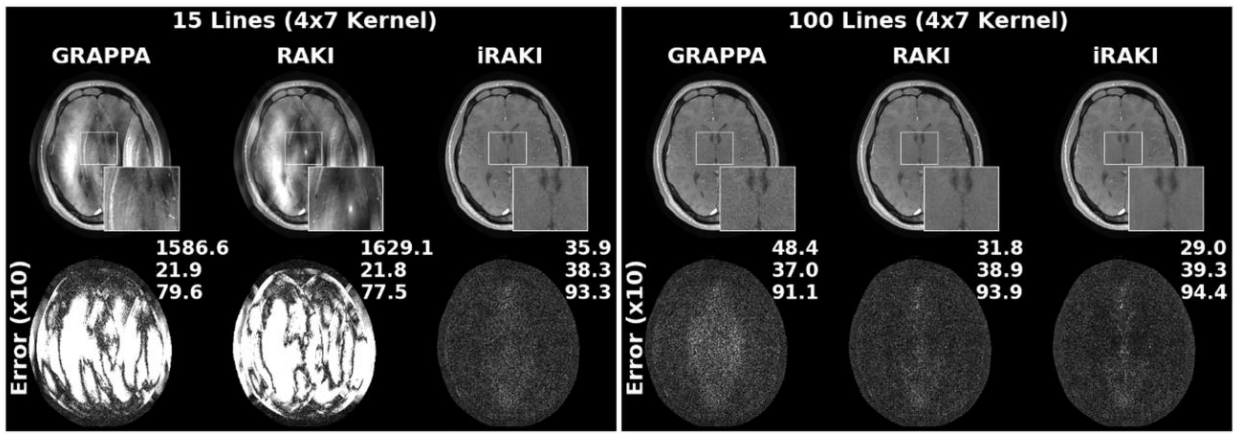

**Figure 3:** GRAPPA, standard RAKI and iRAKI in comparison at 15 and 100 ACS lines as training data (left- and right-hand side, respectively) using a convolution kernel of size 4x7 (Phase- x Readout- direction) assigned to the first convolution layers. Note that the depicted images correspond to samples shown in Figure 2. Error maps with respect to the fully sampled reference image are shown at the bottom including NMSE, PSNR and SSIM difference-metrics.

**In-line calibration**

Figure 4 shows boxplots of the NMSE and SSIM metrics for GRAPPA, RAKI and iRAKI evaluated on fastMRI cohorts with T1, T1post, T2 and FLAIR contrasts. While GRAPPA is generally outperformed for both R=4 and R=5, iRAKI enhances standard RAKI as it yields NMSE medians systematically reduced by 26.4%, 28.7%, 26.2% and 21.7% for T1, T1post, T2 and FLAIR, respectively, for R=4 and 28.3%, 36.4%, 32.6% 30.4% for R=5. Also the SSIM medians are systematically enhanced by 1.5%, 1.4%, 1.7%, 1.6% (R=4) and 2.3%, 2.6%, 3.4%, 3.7% (R=5) for T1, T1post, T2 and FLAIR, respectively.

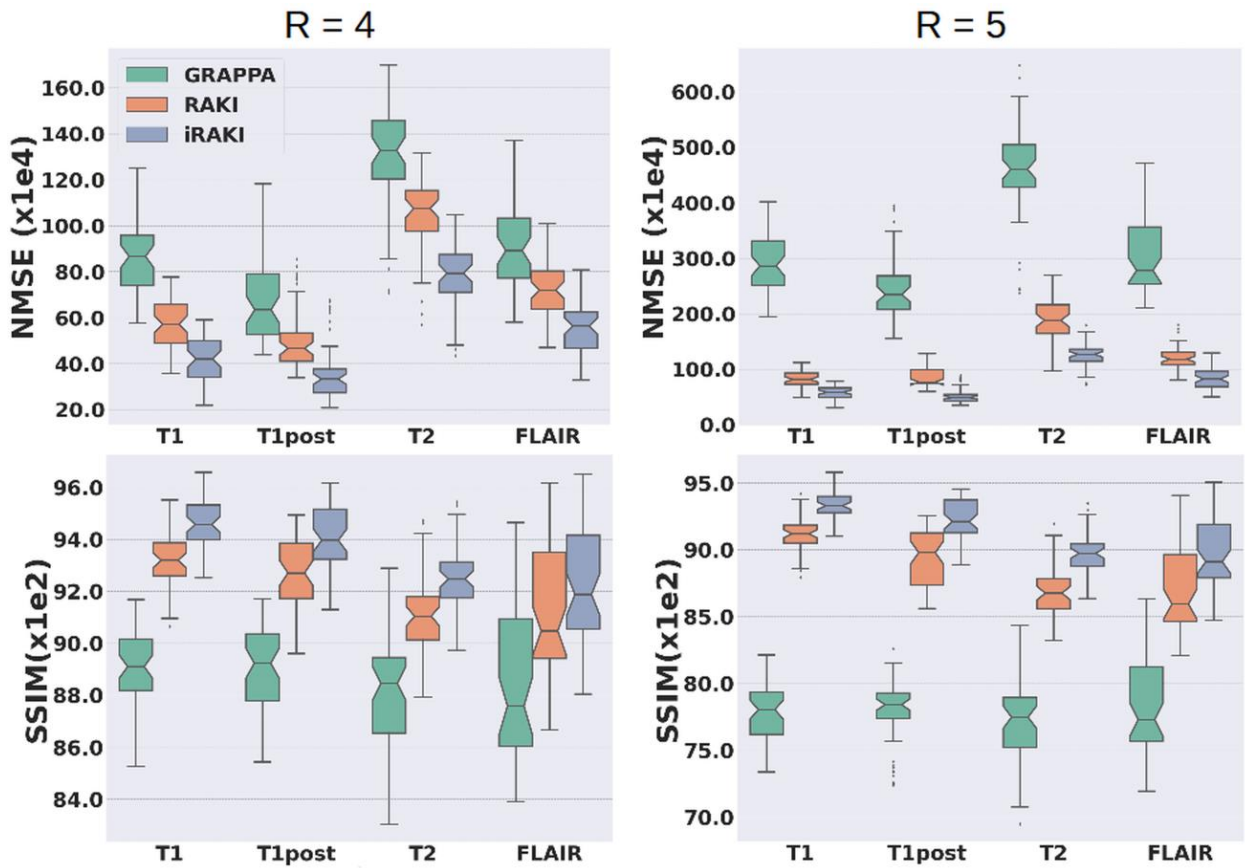

**Figure 4:** Boxplots of NMSE (top) and SSIM (bottom) for GRAPPA, standard RAKI and iRAKI evaluated on cohorts assembled from the fastMRI neuro database. Four different contrast were considered (T1, T1post, T2 and FLAIR), and 50 datasets per contrast were retrospectively undersampled at rate 4 (left) and rate 5 (right). Exemplary image reconstructions are depicted in Figure 5 (R=4) and Figure 6 (R=5).

Figure 5 depicts exemplary GRAPPA, standard RAKI and iRAKI image reconstructions for all evaluated contrasts from the fastMRI cohorts at 4-fold retrospectively undersampling and 18 ACS lines (see Figure 6 for 5-fold retrospectively undersampling and 22 ACS lines). We note that GRAPPA, as expected, suffers from pronounced noise enhancement. Standard RAKI provides noise-resilience, however, suffers from residual artifacts due to limited training data amount for both acceleration factors. iRAKI, however, incorporates desirable features of both GRAPPA and standard RAKI by suppressing noise enhancement and residual artifacts, respectively, resulting in improved visual appearance and outperforming NMSE, PSNR and SSIM for all depicted examples.

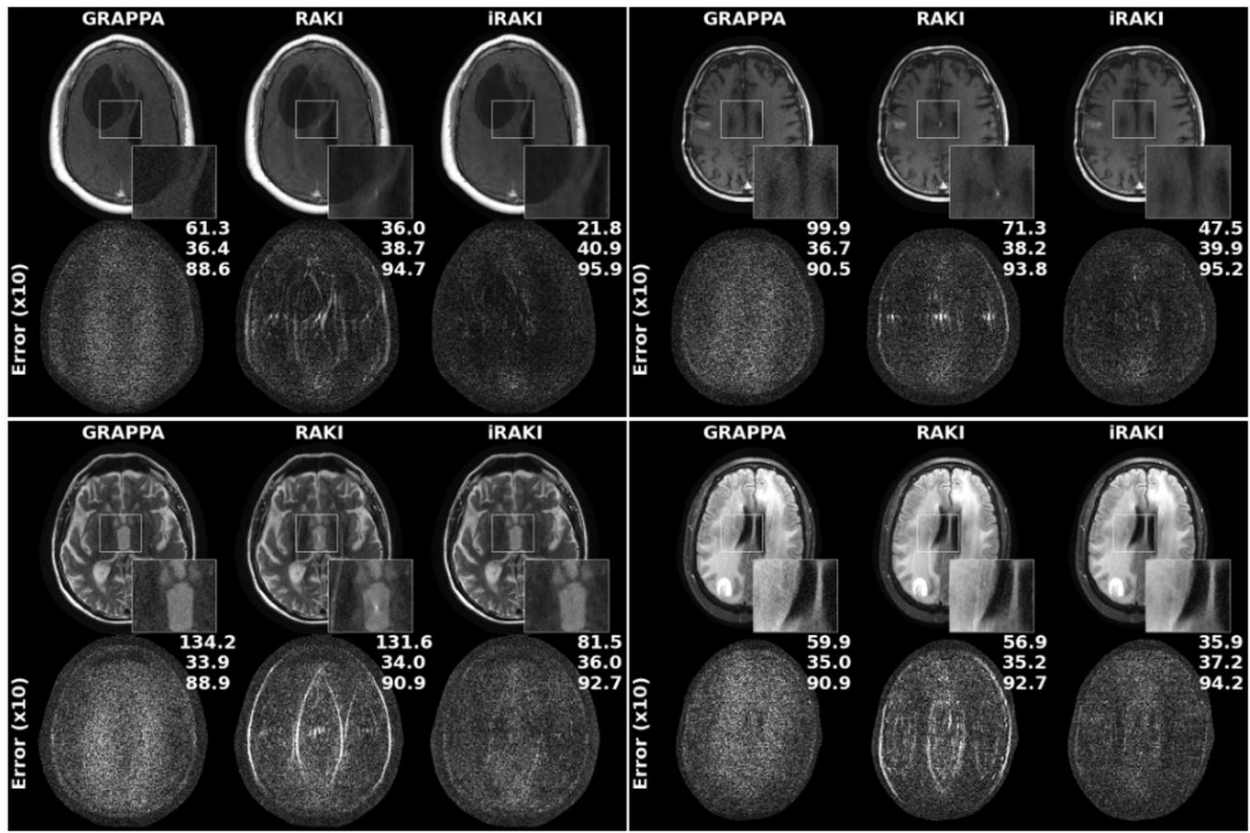

**Figure 5:** GRAPPA, standard RAKI and iRAKI reconstructions for exemplary sample datasets from evaluated fastMRI cohorts (see Figure 4, from left to right, and top to bottom: T1, T1post, T2 and FLAIR). Error maps are shown at the bottom and include NMSE, PSNR and SSIM w.r.t. the fully sampled reference image. Datasets were 4-fold retrospectively undersampled, and 18 ACS lines were used as training data (re-inserted into reconstructed k-space).

Similar outcomes are obtained from three acquired in-plane imaging experiments neuro1-3, for both T1-and T2-weighting and R=4 and R=5 (see supporting information Figure S3-S7, respectively, for image reconstructions with error maps and numeric difference metrics).

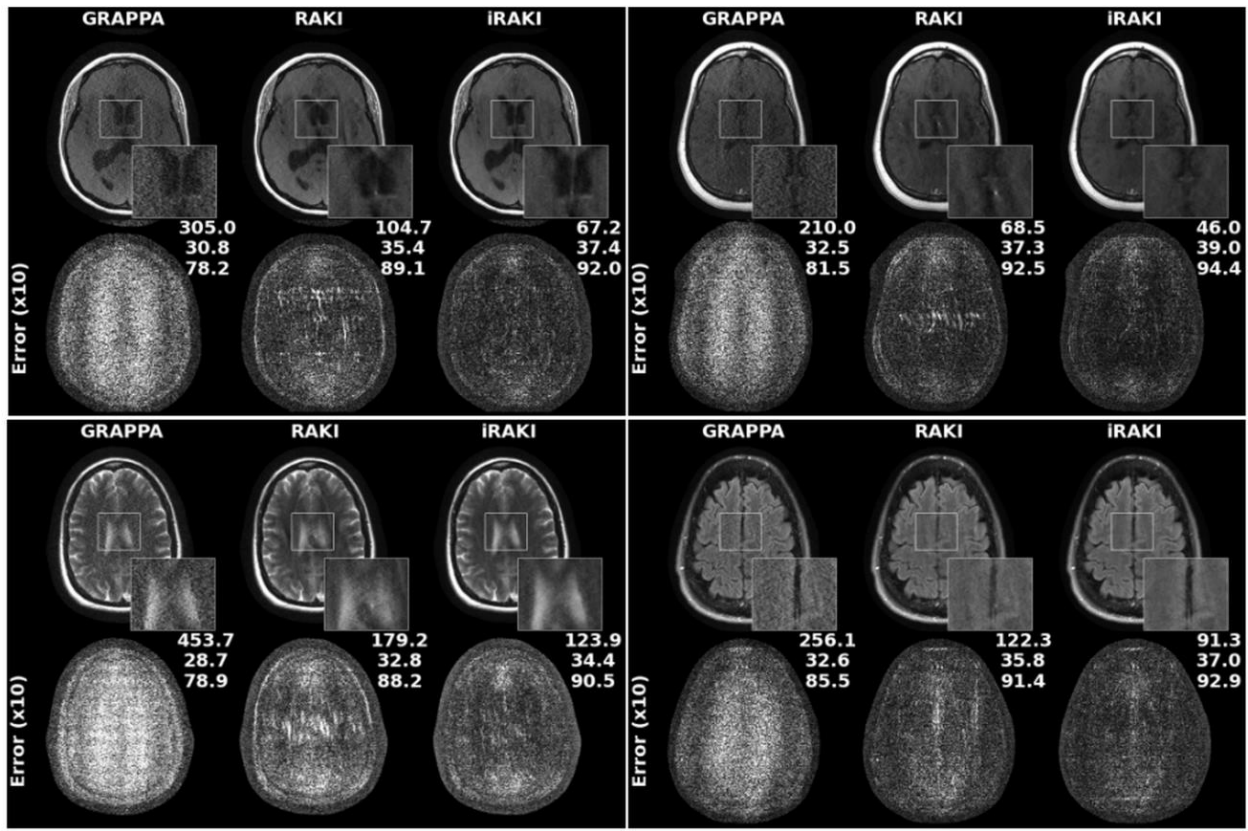

**Figure 6:** GRAPPA, standard RAKI and iRAKI reconstructions for exemplary sample datasets from evaluated fastMRI cohorts (see Figure 4, from left to right, and top to bottom: T1, T1post, T2 and FLAIR). Error maps are shown at the bottom and include NMSE, PSNR and SSIM w.r.t. the fully sampled reference image. Datasets were 5-fold retrospectively undersampled, and 22 ACS lines were used as training data (re-inserted into reconstructed k-space).

**Pre-Scan calibration**

Figure 7 depicts results of GRAPPA, standard RAKI and iRAKI using the pre-scan as training data to reconstruct the 4-fold accelerated neuro4 dataset. Note that the contrast information in the pre-scan (proton-density) varies from that of the undersampled image scan (T1). We observe that standard RAKI is deteriorated due to contrast-loss artefacts (contrast of ACS data sneaks into reconstruction), however, it yields less noise-amplification in comparison to GRAPPA, which maintains contrast. On the other hand, iRAKI does not reveal 'contrast-loss' artifacts appearing in standard RAKI, but preserves its improved noise resilience in comparison to GRAPPA, without the cost of blurring artefacts. We emphasize that iRAKI combines advantages of both GRAPPA and RAKI (more natural contrast and stronger noise suppression, respectively), thus, providing an improved visual appearance.

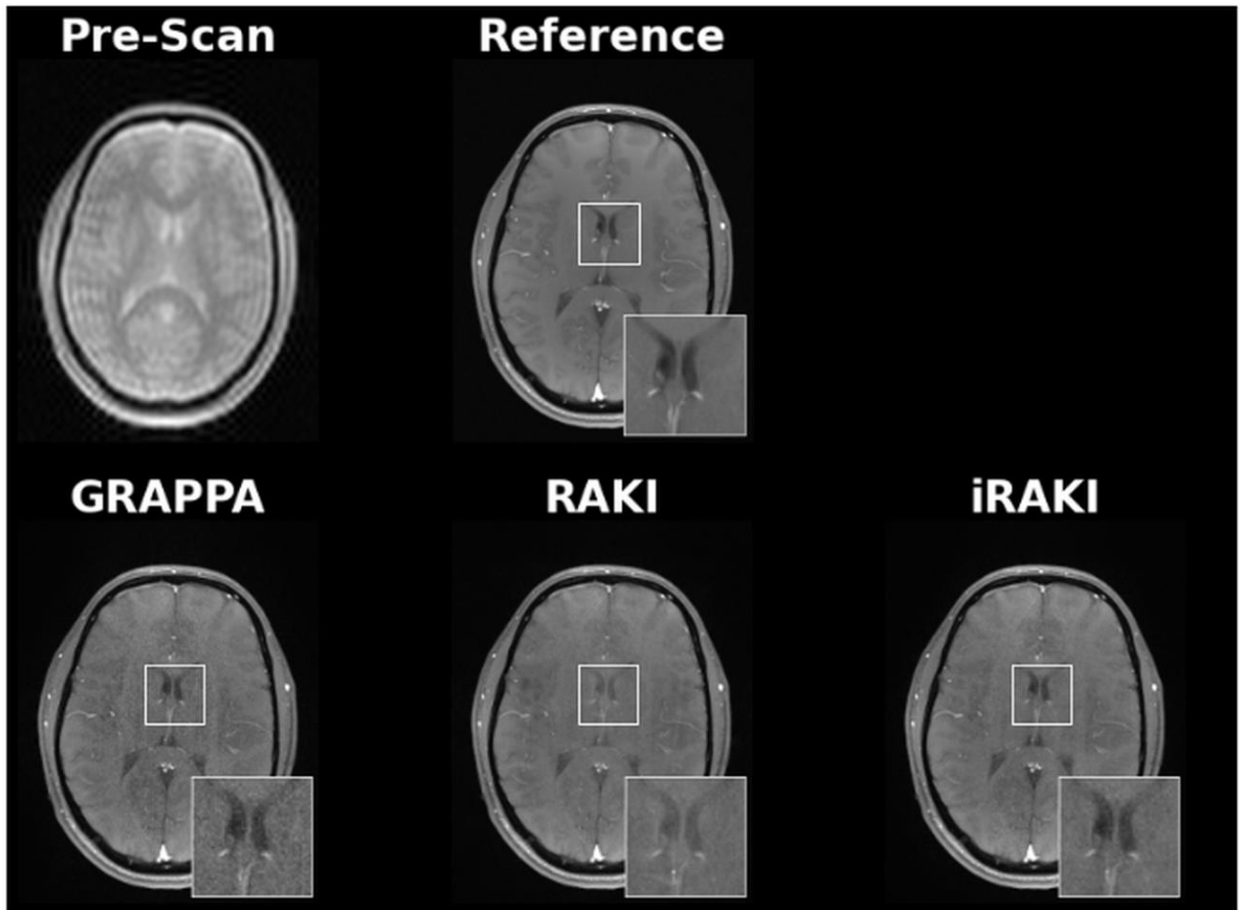

**Figure 7:** A pre-scan (top row) served as training data for image reconstructions of prospectively 4-fold undersampled image scan (neuro4) via GRAPPA, standard RAKI and iRAKI (bottom row).

**Phase-constrained reconstruction**

Figure 8 depicts comparisons of standard iRAKI with iRAKI including phase constraints via the VCC concept (iRAKI-VCC) using the T1-neuro1 and the T2-neuro3 datasets. For both 4- and 5-fold undersampling using 18 and 22 original ACS lines, respectively, iRAKI-VCC enhances standard iRAKI as it yields improved suppression of both residual artifacts and noise-enhancement, leading to greatly improved visual appearance and improved NMSE, PSNR and SSIM.

Note that comparisons of GRAPPA, standard RAKI and iRAKI with and without VCC included for all three datasets neuro1-neuro3 are depicted in the supporting information Figures S3-S7, respectively. We underline that standard RAKI-VCC provides an improved noise resilience in comparison to GRAPPA-VCC, however, it still reveals residual artifacts, which are not apparent in iRAKI-VCC.

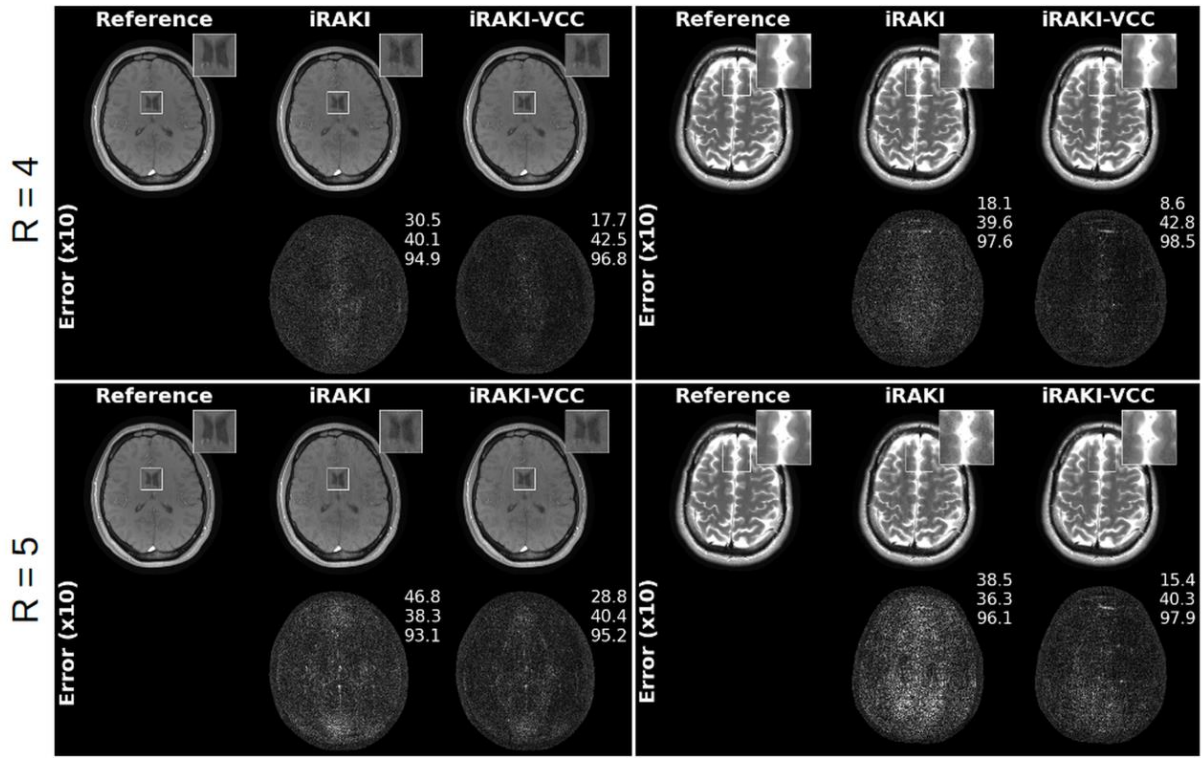

**Figure 8**: Standard iRAKI in comparison to phase-constrained iRAKI via the Virtual-Conjugate-Coils concept (iRAKI-VCC) at 4- and 5-fold retrospective undersampling (top and bottom row, respectively), evaluated on the T1 neuro2- (left column) and T2 neuro3 datasets (right column).

## Comparison to image-based end-to-end Variational Network

Figure 9 A depicts a 4-fold retrospectively undersampled T1 knee dataset reconstructed using scan-specific iRAKI, and a variational network (VarNet). The VarNet was trained on a database of knee datasets with matching contrast. While both iRAKI and the VarNet provide high quality reconstructions, the VarNet shows improved denosing effects, as it reveals a better noise resilience in comparison to iRAKI, also indicated by improved quantitative metrics.

However, using the same VarNet in order to reconstruct a 4-fold retrospectively undersampled dataset from the T1post fastMRI neuro database, which represents a cross-domain task regarding anatomy and contrast of the target image, we occasionally observe the emerge of severe residual artifacts, as exemplary depicted in Figure 9 B. On the other hand, scan-specific iRAKI generally retains its improved reconstruction quality, and shows high flexibility across different anatomy- or contrast information. The above mentioned findings were also observed for datasets of the fastMRI cohorts with T1-, T2- and FLAIR-weighting (see supporting information Figure S8 for exemplary image reconstructions).

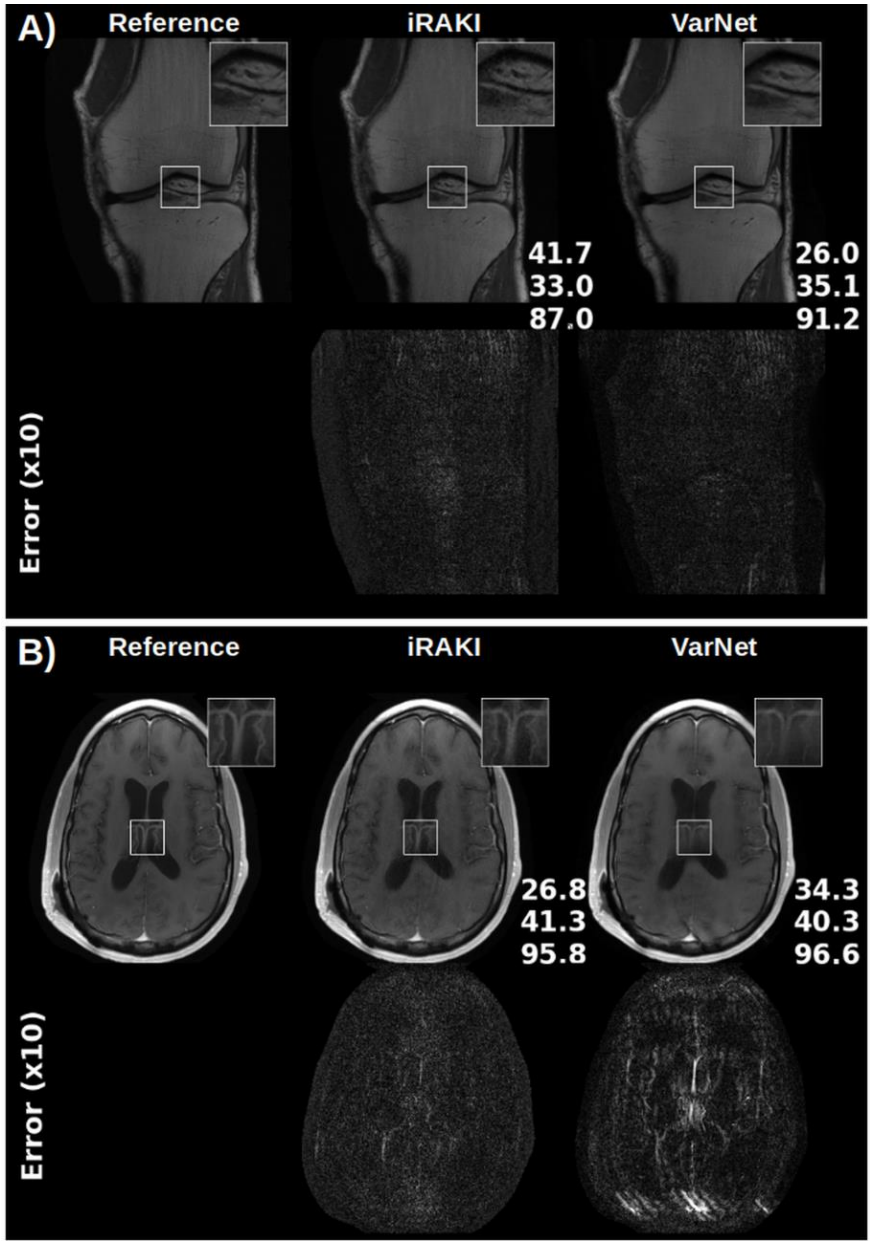

**Figure 9:** Comparison between iRAKI and variational network (VarNet) **A)** with matching anatomy and contrast in training- and reconstructed data and **B)** with non-matching anatomy and contrast.

## 4. Discussion

This study aimed to enhance the deep-learning method RAKI in the limit of only few ACS lines in standard 2D imaging at medium acceleration factors. To this end, an optimized RAKI model was evaluated in a first step. Compared to the original RAKI model, it achieved significantly lower NMSE compared to the original RAKI model on test data from the fastMRI neuro database. However, the performance of optimized RAKI significantly declines when the number of ACS lines is further decreased. An iterative training approach is proposed that relies on an initial GRAPPA reconstruction for training data augmentation, and iterative refining of the CNN weights using original and augmented ACS. To evaluate its robustness, iRAKI was tested on 200 different datasets assembled from the fastMRI neuro database. Using only 18 and 22 ACS lines for R=4 and R=5, respectively, it yields systematically enhanced NMSE and SSIM for different contrast settings in comparison to GRAPPA and standard RAKI. On a visual scale, iRAKI suppresses residual artifacts apparent in standard RAKI, while showing less noise-enhancement than GRAPPA. Thus, iRAKI incorporates beneficial features from both methods.

The improved performance of iRAKI can be attributed to several aspects. First, the augmented ACS data allows for the use of a larger convolution filter size, and helps to better capture the extended k-space "footprint" of the coil sensitivities(9)(10)(22). Second, the augmented ACS data gain in accuracy after each reconstruction step. Analogous to SPIRiT (31), both original and reconstructed data are used during k-space interpolation. After each iteration, the reconstructed samples become more accurate after applying the reconstruction kernel and re-inserting the original measured data. We emphasize that in iRAKI the original ACS lines are re-inserted after each iteration step, and the augmented ACS provide additional information. In contrast to SPIRiT, the convolution weights are refined after each iteration in iRAKI. Decreasing the initial learning rate after each iteration ensures further robustness.

Although GRAPPA can further be improved by iterative training, its performance is limited by the linearity. As shown in supporting information Figure S 9, iRAKI outperformed iterative-GRAPPA for a wide range of imaging scenarios.

iRAKI relies on an initial GRAPPA reconstruction for training data augmentation. It is worth noting that this approach has also been proposed, but not demonstrated in the original RAKI article (4). An initial parallel imaging reconstruction is also used in Scan-Specific Artifact Reduction in k-Space (SPARK) (32), where a CNN is used to correct for k-space artifacts of the initial reconstruction to achieve better reconstruction quality. Future investigations may focus on the evaluation of an initial iRAKI reconstruction in SPARK to further enhance reconstruction quality.

Another approach based on an initial fast reconstruction is inspired by Low-Rank Matrix Modelling of Local k-Space Neighborhoods (LORAKS) (33) and RAKI, and is termed LORAKI (34). LORAKI translates the linear Auto-Calibrated-LORAKS method into a nonlinear deep learning method. LORAKI admits a wide range of sampling patterns, and even calibrationless patterns are possible if synthetic ACS data is generated with a fast initial reconstruction. However, it requires tuning of multiple parameters (e.g. rank, kernel sizes, regularization parameters) and needs VCCs to capture the LORAKS phase constraints. In contrast, iRAKI is more flexible in terms of phase-constraints, and the use of VCCs is optional. It is worth noting that k-space inconsistencies such as non-periodic flow or motion may prevent the use of phase-constraints.

Additional flexibility of iRAKI stands in the varying contrast information between calibration- and undersampled data, as it prevents contrast contamination in standard RAKI (also shown in (35)), while preserving its noise-resilience. In this work it was demonstrated that iRAKI provides better image quality than GRAPPA and standard RAKI for the case of pre-scan calibration, that is often used in parallel imaging.

**Limitations and Outlook**

The performance of iRAKI was evaluated for standard 2D imaging so far. Its applicability for 3D imaging (36)(37), wave-encoding (38), or simultaneous multi-contrast reconstruction (JVC-GRAPPA) (38) needs to be investigated in future works.

One drawback of iRAKI is the increased reconstruction time. The total training time in iRAKI amounts to $\approx$ 180 s and $\approx$ 170 s for $R = 4$ and $R = 5$, respectively, exceeding standard RAKI ($\approx$ 12s and $\approx$ 20s) and GRAPPA ($<$ 1.5 s both $R = 4$ and $R = 5$). Future work should focus on optimizing the training speed to apply iRAKI in clinical applications.

The performance of iRAKI is expected to depend on the reconstruction quality of the initial GRAPPA reconstruction. A general acceleration factor limit for the application of iRAKI cannot be specified, as the quality of the initial GRAPPA reconstruction depends on the base signal-to-noise ratio and the g-factor (2)(40). In our experience, iRAKI yields better image quality than GRAPPA and standard RAKI at R=6, but the reconstructed images may not be of diagnostic value because of residual artifacts and noise enhancement (see supporting information Figure S10). However, one should keep in mind that the acquisition of many ACS lines comes along with decreases overall acceleration. Depending on the matrix size of the final image, it may be faster to scan with 4-fold undersampling and 18 ACS lines instead of 6-fold undersampling and 48 ACS lines.

In this work, it was also investigated whether iRAKI yields comparable results to the image-based variational networks. The variational network showed better denoising performance in comparison to iRAKI, when it is trained on a database, whose anatomy and contrast information matches to the target undersampled image. However, as shown in this work, variational networks may suffer from residual artifacts when this condition is not fulfilled, in accordance to previous findings (41). In case of the development of novel MR sequences or the examination of uncommon anatomies, there may be a lack of large databases for end-to-end network training. For these situations, scan-specific iRAKI may be beneficial as it does not require large databases.

## 5. Conclusion

The number and contrast of training samples are essential for standard RAKI reconstruction quality. Given a limited training data amount, the proposed scan-specific iRAKI combines beneficial features of GRAPPA and standard RAKI and yields reconstructions with suppression of both noise and residual artefacts for standard 2D imaging. It shows flexibility in terms of different calibration approaches (pre-scan or integrated ACS) and implementation of phase-constrained reconstruction.

## Acknowledgments

The authors would like to thank Moritz Blumenthal for implementing variational networks in BART and Dr. Fabian T. Gutjahr for constructive criticism of the manuscript. We thank dataSphere@JMU for providing informative discussions and the German Federal Ministry of Education and Research (BMBF) for funding project line VIP+ (03VP04951).

# Supporting Information

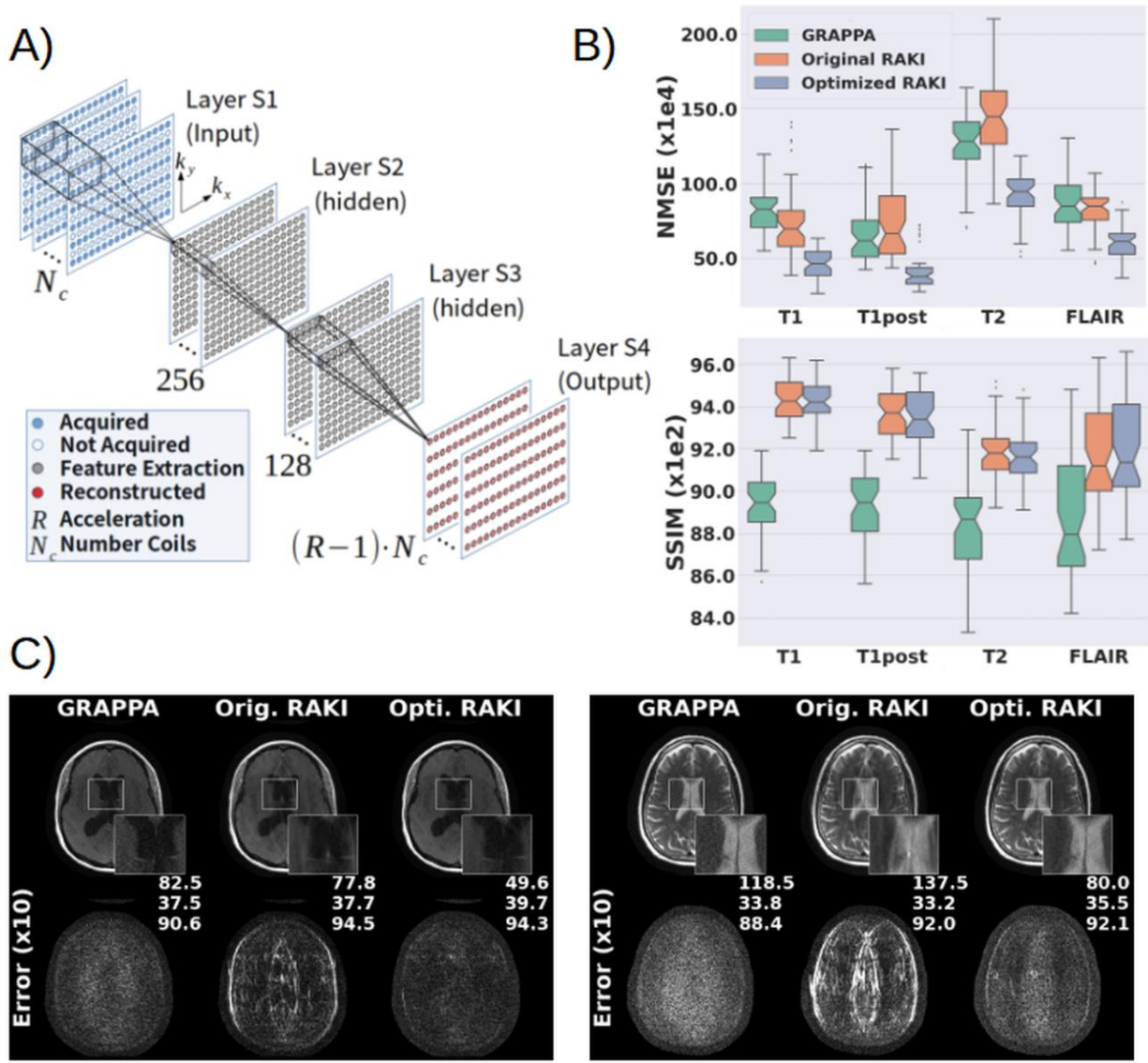

**Figure S1**

A) Architecture of the optimized Convolutional Neural Network (CNN) used for RAKI implementation in this work. Convolution is performed with complex-valued filter matrices. The input-layer takes in the zero filled, k-space data, thus having $N_c$ channels, with $N_c$ denoting the number of receiver coils. The first and second hidden layer are assigned 256 and 128 channels, respectively. The output layer predicts all missing k-space data across all coils, thus having $(R - 1) \cdot N_c$ channels, with $R$ denoting the undersampling rate.

B) Boxplots of the normalized mean-squared-error (NMSE, top) and the structural similarity index measure (SSIM, bottom) for GRAPPA, original RAKI and optimized RAKI evaluated on cohorts from the fastMRI neuro database (50 datasets for each T1, T1post, T2 and FLAIR).

C) GRAPPA, original RAKI and optimized RAKI in comparison shown for exemplary sampled datasets of evaluated fastMRI cohorts from B. Error maps are shown below, and include NMSE, PSNR and SSIM (from top to bottom).

**Supporting Material S2:** mp4-movie

Image reconstructions of GRAPPA, standard RAKI and iRAKI evaluated and varying training data amount (see Figure 2). Error maps are shown below, and include NMSE, PSNR and SSIM difference metrics w.r.t. the fully sampled reference image.

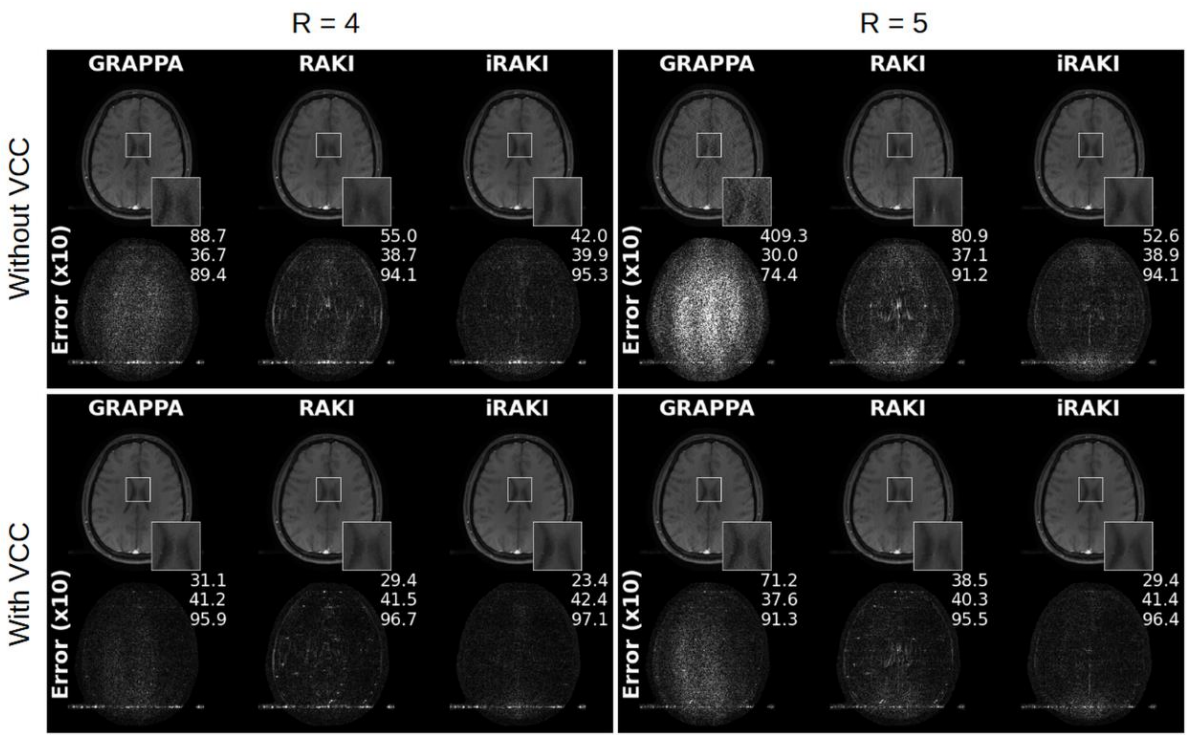

**Figure S3:** GRAPPA, standard RAKI and iRAKI evaluated on the T1-neuro2 dataset at 4- and 5-fold retrospectively undersampling (R=4, left column and R=5, right column). Error maps are shown below, and include NMSE, PSNR and SSIM. Reconstructions were performed without phase constrains (top row) and including the virtual-conjugate-coils (VCC) concept (bottom row).

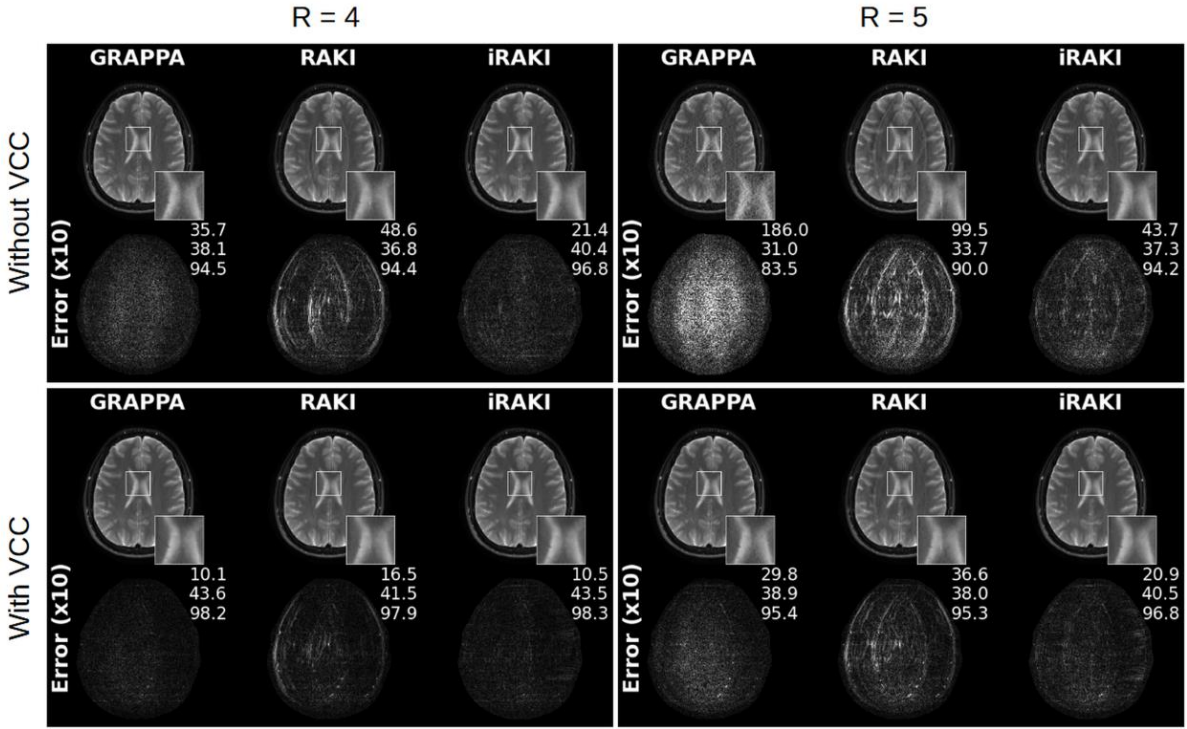

**Figure S4:** GRAPPA, standard RAKI and iRAKI evaluated on the T2-neuro2 dataset at 4- and 5-fold retrospectively undersampling (R=4, left column and R=5, right column). Error maps are shown below, and include NMSE, PSNR and SSIM. Reconstructions were performed without phase constrains (top row) and including the virtual-conjugate-coils (VCC) concept (bottom row).

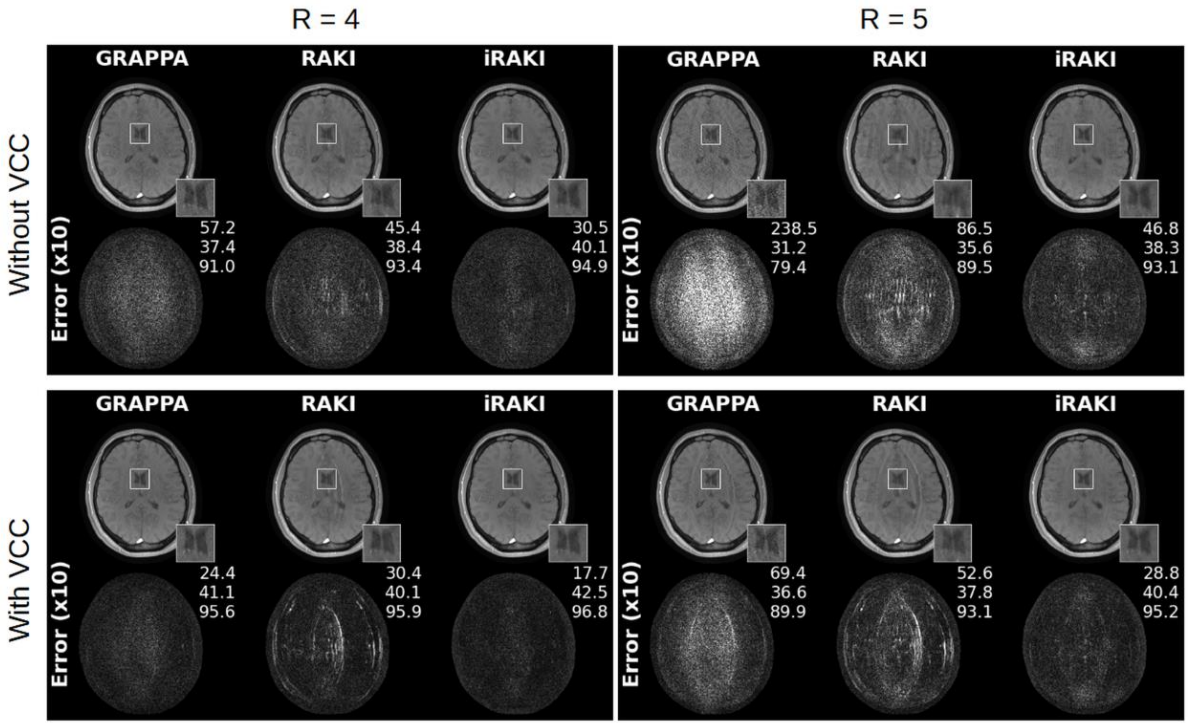

**Figure S5:** GRAPPA, standard RAKI and iRAKI evaluated on the T1-neuro1 dataset at 4- and 5-fold retrospectively undersampling (R=4, left column and R=5, right column). Error maps are shown below, and include NMSE, PSNR and SSIM. Reconstructions were performed without phase constrains (top row) and including the virtual-conjugate-coils (VCC) concept (bottom row).

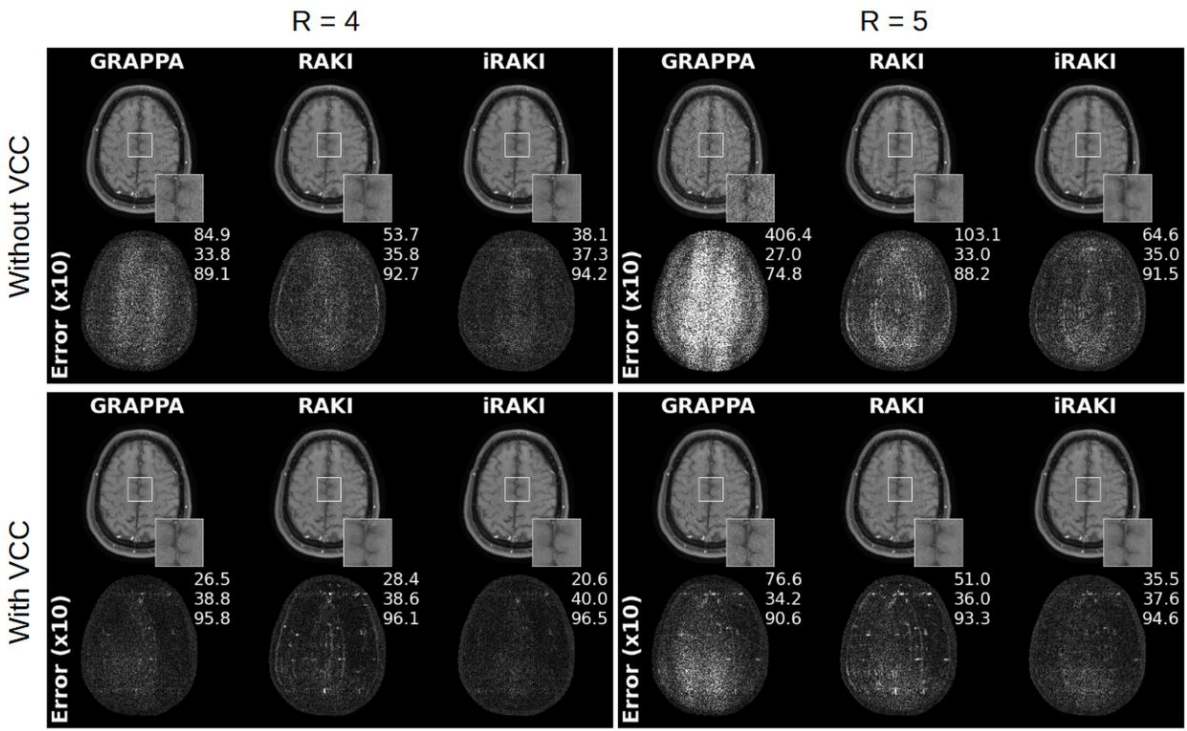

**Figure S6:** GRAPPA, standard RAKI and iRAKI evaluated on the T1-neuro3 dataset at 4- and 5-fold retrospectively undersampling (R=4, left column and R=5, right column). Error maps are shown below, and include NMSE, PSNR and SSIM. Reconstructions were performed without phase constrains (top row) and including the virtual-conjugate-coils (VCC) concept (bottom row).

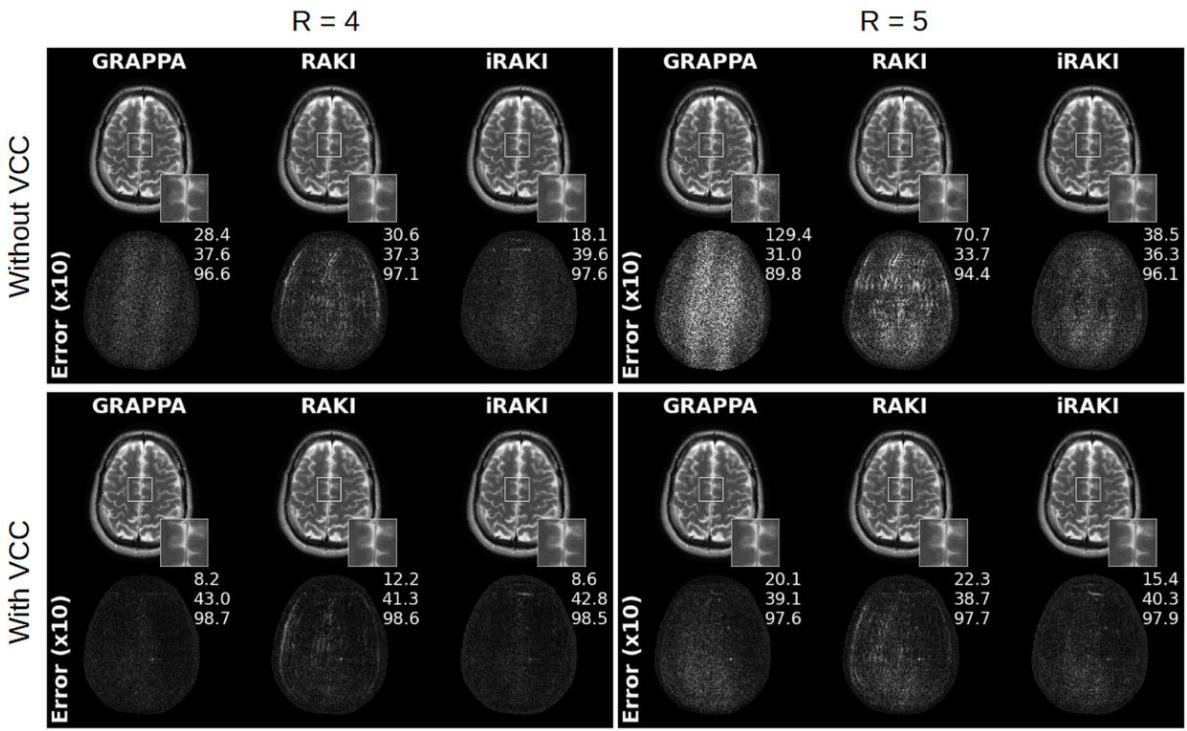

**Figure S7:** GRAPPA, standard RAKI and iRAKI evaluated on the T2-neuro3 dataset at 4- and 5-fold retrospectively undersampling (R=4, left column and R=5, right column). Error maps are shown below, and include NMSE, PSNR and SSIM. Reconstructions were performed without phase constrains (top row) and including the virtual-conjugate-coils (VCC) concept (bottom row).

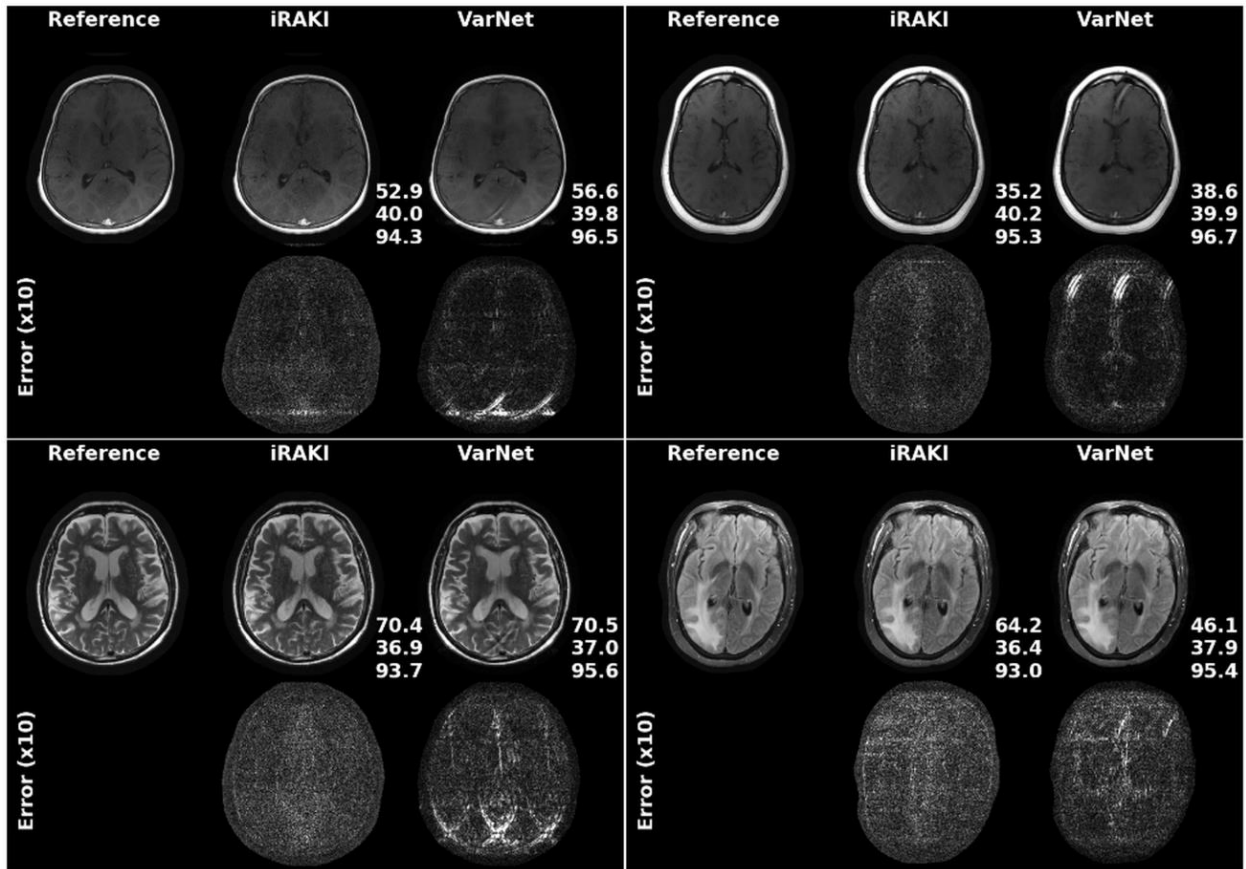

**Figure S8**: iRAKI and the variational network (VarNet) in comparison for scans with

T1-,T1post-, T2- and FLAIR weighting (from left to right and top to bottom). The VarNet was trained one knee-data with proton-density weighting.

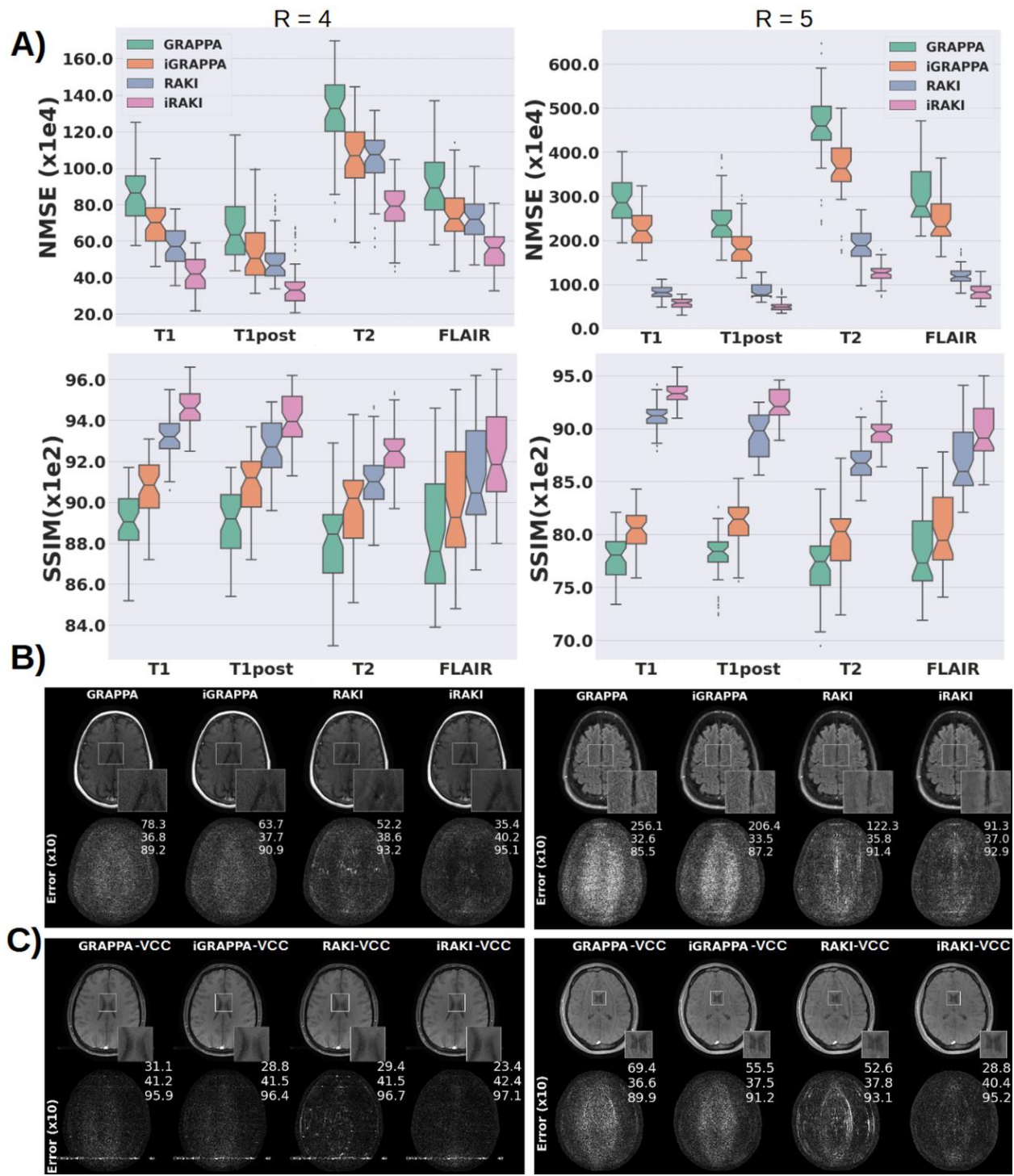

**Figure S9**

A) Boxplots of NMSE (top) and SSIM (bottom) for GRAPPA, iterative-GRAPPA, standard RAKI and iRAKI evaluated on cohorts assembled form the fastMRI neuro-database. Four different contrast were considered (T1, T1post, T2 and FLAIR), and 50 datasets per contrast were retrospectively undersampled at rate 4 (left) and rate 5 (right) using 18 and 22 ACS lines, respectively.

B) Exemplary image reconstructions from evaluations depicted in A) (left: T1post 4-fold undersampling, right: FLAIR 5-fold undersampling).

C) T1-neuro2-dataset (left, 4-fold undersampling, 18 ACS lines) and T1-neuro1-dataset (right, 5-fold underampling, 22 ACS lines) reconstructed with GRAPPA, iterative-GRAPPA, standard RAKI and iRAKI including the VCC concept.

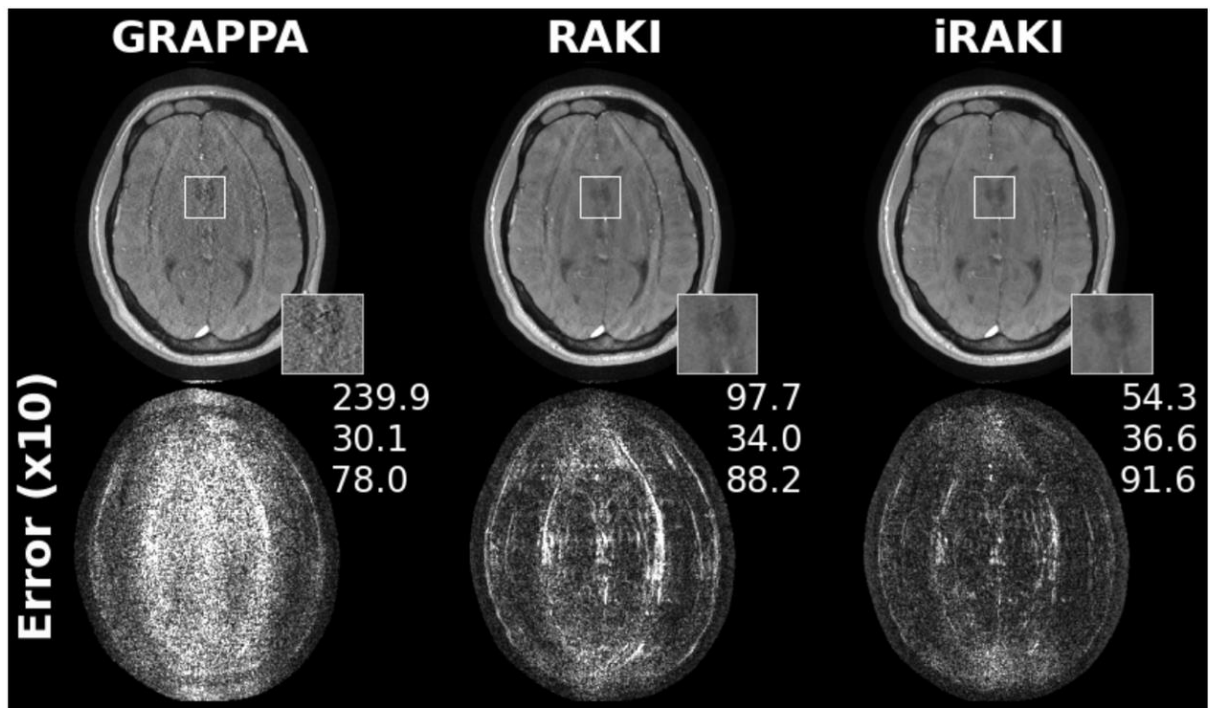

**Figure S10:** GRAPPA, standard RAKI and iRAKI evaluated on 6-fold retrospectively undersampling of T1-neuro1-dataset. 28 ACS lines were used as training data (re-inserted into reconstructed k-spaces). Error maps are shown below, and include NMSE, PSNR and SSIM difference metrics.